\documentclass[aps,pra,twocolumn]{revtex4}
\usepackage{epsfig,dsfont,amssymb,amsmath,amsthm,amsfonts,amsbsy,mathrsfs}
\usepackage{graphicx}
\usepackage{bm}%bold math
\begin{document}
\title{Cross-Kerr nonlinearity between continuous-mode coherent states and single photons}
\author{Bing He$^1$, Qing Lin$^2$, and Christoph Simon$^1$}
\affiliation{$^1$ Institute for Quantum Information Science
and Department of Physics and Astronomy, University of
Calgary, Calgary T2N 1N4, Alberta, Canada\\
$^2$ College of Information Science and Engineering,
Huaqiao University (Xiamen), Xiamen 361021, China}

\begin{abstract}
Weak cross-Kerr nonlinearities between single photons and coherent states are the basis for
many applications in quantum information processing. These nonlinearities have so far mainly
been discussed in terms of highly idealized single-mode models. We develop a general theory of the interaction between continuous-mode photonic pulses and apply it to the case of a single photon interacting with a coherent state. We quantitatively study the validity of the usual single-mode approximation using the concepts of fidelity and conditional phase. We show that
high fidelities, non-zero conditional phases and high photon numbers are compatible, under conditions where the pulses fully pass through each other and where unwanted transverse-mode effects are suppressed.
\end{abstract}

\maketitle

\section{introduction}
Photons are among the main candidate systems for the implementation of quantum information processing. One potential avenue in this context is to implement quantum gates between individual photonic qubits. Although there has been considerable experimental progress in realizing photonic pulse cross-phase modulation (XPM) in recent years \cite{zhu, chen, mat, Lo}, this is still an extremely challenging goal, because the interaction between single photons in nonlinear media is generally weak.

An alternative attractive approach involves the interaction of individual photons with intense coherent states. This approach can be used for the implementation of nondemolition measurements of the photon number \cite{imoto,munroND}. It also forms the basis for quantum gate proposals between individual photons that use the coherent state as an auxiliary system \cite{n-m-04, b-s-05, m-05}. For all of these applications, given an idealized single-mode coherent state $|\alpha\rangle$ and a single photon state $|1\rangle$, one aims to perform the transformation $|\alpha\rangle|1\rangle\rightarrow |\alpha e^{i\theta_c}\rangle|1\rangle$, where $\theta_c$ can be small. It was first proposed in \cite{n-m-04, b-s-05} that a deterministic parity gate for single-photon qubits could be realized by such weak nonlinearity. This idea has been widely employed in the researches on physical realization of quantum communication and quantum computing, see, e.g., \cite{j-05, k-p-07, h-b, t-k, s-d, l-l,l-h, n-k-k} for recent studies.

Most of the previous studies adopt a single mode assumption for XPM. Under this assumption the interaction term for the photonic states is given as $\hat{V}=-\chi \hat{a}^{\dagger}_s \hat{a}_s\hat{a}^{\dagger}_c \hat{a}_c$, where $\hat{a}_s$ and $\hat{a}_c$ represent the mode
of single photon and coherent state, respectively, and $\chi$ is the interaction strength ($\hbar=1$ is used throughout this paper). This actually models an ideal XPM mentioned above as the coupling of two harmonic oscillators, one of which is in the state $\hat{a}^{\dagger}_s |0\rangle$ and the other is in the state $\exp(\alpha \hat{a}_c^{\dagger}-\alpha^{\ast}\hat{a}_c)|0\rangle$, through the interaction described by the nonlinear term $\hat{V}$.

This type of single-mode description is highly idealized. In reality, even if the light pulses start out as single-mode, interactions will generically create continuous-mode entanglement.
This has been analyzed for photon-photon gates, studying both the longitudinal \cite{eit-review,lukin-imamoglu,petrosyan} and, more recently, the transverse \cite{trans} degrees of freedom. However, to our knowledge, no such study has been made for the equally important case of single photons interacting with coherent states.

Here we develop an effective quantum field theory description of the XPM between continuous-mode photonic pulses. The photonic degrees of freedom are described by quantum fields over the whole space that interact with each other through a general potential $\Delta({\bf x}-{\bf x'})$. The interaction is thus treated as instantaneous, which corresponds to an adiabatic elimination of the degrees of freedom of the nonlinear medium. This approximation is well established for nonlinearities based on electromagnetically induced transparency (EIT) \cite{eit-review, lukin-imamoglu, petrosyan,xpm}. It has also been used to describe interactions between Rydberg polaritons \cite{headon-1}, and photonic nonlinearities due to collisions in Bose-Einstein condensates (BECs) \cite{rispe}. We will furthermore focus on very short-range interactions, which are typically modeled by the contact potential $\Delta({\bf x}-{\bf x'})=\eta\delta({\bf x}-{\bf x'})$. For a discussion of non-instantaneous effects see Ref. \cite{sh-07}.

This paper is organized as follows. In Sec. II we discuss how the quantum states of continuous-mode pulses can be expressed in terms of quantum field operators. In Sec. III we describe their interactions with the help of the above-mentioned interaction potential approach. In Sec. IV we introduce an interaction picture for the interacting quantum field model. In Sec. V we apply the developed formalism to several relevant cases. In particular in Sec. V.A we study the case of two interacting single photons, introducing the concepts of fidelity and conditional phase. In Sec. V.B we treat the case of a coherent state interacting with a single photon, adapting the aforementioned concepts. In Sec. VI we study the performance of cross-Kerr nonlinearity based on pulses interacting through contact potential. In particular, we discuss the the different types of XPM with pulse of unequal and equal group velocities. Finally in section VII we give our conclusions.

\section{Continuous-mode photonic states}

We first clarify the quantum states for the realistic pulses. Any classic electric field
${\bf E}$ (magnetic field ${\bf H}$) can be expressed in terms of the plane wave expansion
\begin{eqnarray}
{\bf E}({\bf x},t)&=&{\bf E}^{+}({\bf x},t)+{\bf E}^{-}({\bf x},t)\nonumber\\
&=&\sum_{{\bf k},\lambda}\hat{\epsilon}_{{\bf k},\lambda}\sqrt{\frac{  \omega_{\bf k}}{2\epsilon_0 V}}c_{{\bf k},\lambda}e^{-i\omega_{\bf k}t+i{\bf k}\cdot {\bf x}}+C.~c,
\end{eqnarray}
where $\hat{\epsilon}_{{\bf k},\lambda}$ with $\lambda=1,2$ represent the polarization vectors, and $C.~c.$ stands for the complex conjugate. If the spatial volume $V$ tends to infinity, the discrete sum with respect to ${\bf k}$ will be replaced by the integrals over the continuous spectrum. The quantization of the field is straightforwardly performed by replacing the amplitudes $c_{{\bf k},\lambda}$ with the annihilation operators $\hat{a}_{{\bf k},\lambda}$ \cite{s-z}:
\begin{eqnarray}
{\hat{\bf E}({\bf x},t)}&=&\hat{\bf E}^{+}({\bf x},t)+\hat{\bf E}^{-}({\bf x},t)\nonumber\\
&=&\sum_{{\bf k},\lambda}\hat{\epsilon}_{{\bf k},\lambda}\sqrt{\frac{  \omega_{\bf k}}{2\epsilon_0 V}}\hat{a}_{{\bf k},\lambda}e^{-i\omega_{\bf k}t+i{\bf k}\cdot {\bf x}}+H.~c.,
\label{quant}
\end{eqnarray}
where $H.c.$ stands for the Hermitian conjugate. A current distribution ${\bf J}({\bf x},t)$ acted by a field with the electromagnetic potential ${\bf A}({\bf x},t)$, where ${\bf E}({\bf x},t)=-\partial_t{\bf A}({\bf x},t)$ and ${\bf H}({\bf x},t)=\nabla\times {\bf A}({\bf x},t)$, will radiate an electromagnetic field in coherent state with multiple modes \cite{s-z}:
\begin{eqnarray}
|\{\alpha_{\bf k}\}\rangle&=& \exp \{-i\int_0^t dt'\hat{V_c}(t')\}|0\rangle\nonumber\\
&=&\prod_{\bf k}\exp(\alpha_{\bf k}\hat{a}_{\bf k}^{\dagger}-\alpha_{\bf k}^{\ast}\hat{a}_{\bf k})|0\rangle=\prod_{\bf k}|\alpha_{\bf k}\rangle,
\label{m-coherent}
\end{eqnarray}
where $\hat {V_c}(t)=\int d^3x~ {\bf J}({\bf x},t)\cdot {\bf A}({\bf x},t)$ and
\begin{eqnarray}
\alpha_{\bf k}= \sqrt{\frac{1}{2\epsilon_0 \omega_{\bf k} V}}\int_0^t dt'\int d^3x \hat{\epsilon}_{{\bf k}}\cdot {\bf J}({\bf x},t) e^{i\omega_{\bf k}t'-i{\bf k}\cdot {\bf x}}.
\end{eqnarray}
A pulse generated in the above process carries a continuous spectrum of the modes ${\bf k}$.

We here adopt a systematic way to express the states of photonic pulses in terms of
the polarization components of a slowly varying field $\hat{\bf E}^{+}({\bf x},t)$.
For any field sharply peaked around a certain frequency, the slowly varying frequency $\omega_{\bf k}$ in the square-root factor of (\ref{quant}) can be regarded as constant \cite{s-z}. Then the dynamical behavior of the slowly varying field will be simply reduced to that of the slowly varying envelope $\hat{\Psi}({\bf x},t)$ corresponding to its polarization components \cite{g-c-08}. The detailed relation between the field operators $\hat{\bf E}^{+}({\bf x},t)$
and $\hat{\Psi}({\bf x},t)$ can be found in \cite{g-c-08}. For simplicity we will not distinguish between different polarization components. The field operator $\hat{\Psi}({\bf x},t)$ satisfies the simple equal-time commutation relation \cite{g-c-08}
\begin{eqnarray}
[\hat{\Psi}({\bf x},t), \hat{\Psi}^{\dagger}({\bf x'},t)]=\delta ({\bf x}-{\bf x'}).
\label{field}
\end{eqnarray}

Using the field operator $\hat{\Psi}({\bf x},0)\equiv \hat{\Psi}({\bf x})$, a multi-mode coherent state generated in the process of Eq. (\ref{m-coherent}) can be written as
\begin{eqnarray}
|\{\alpha_{\bf k}\}\rangle=\exp \{ \int d^3 x\alpha({\bf x})\hat{\Psi}_{}^{\dagger}({\bf x})-\int d^3 x\alpha^{\ast}({\bf x})\hat{\Psi}_{}({\bf x})\}|0\rangle.~
\end{eqnarray}
The amplitudes of the modes are the Fourier transforms of $\alpha({\bf x})$:
\begin{eqnarray}
\alpha_{\bf k}= \frac{1}{(2\pi)^{\frac{3}{2}}}\int d^3 x \alpha({\bf x}) e^{-i{\bf k}\cdot {\bf x}},
\end{eqnarray}
where we have considered a continuous pulse spectrum.
Similarly a continuous-mode single photon state $|1\rangle=\sum_{{\bf k}}\xi_{{\bf k}}a^{\dagger}_{\bf{k}}|0\rangle$ can be expressed as
\begin{eqnarray}
|1\rangle= \int d^3 x f({\bf x})\hat{\Psi}^{\dagger}({\bf x})|0\rangle,
\end{eqnarray}
with
\begin{eqnarray}
\xi_{{\bf k}}=\frac{1}{(2\pi)^{\frac{3}{2}}}\int d^3 x f({\bf x})e^{-i{\bf k}\cdot {\bf x}}.
\end{eqnarray}
The function $\alpha({\bf x})$ and $f({\bf x})$ depict the pulse shapes.

\section{dynamics of interacting pulses}
To completely understand the XPM between photonic pulses, it is necessary to find the output state of the pulses from the initial product state, e.g., $|\Psi_{in}\rangle=|\{\alpha_{\bf k}\}\rangle\otimes |1\rangle$. The formalism provided above allows one to obtain the output state from the dynamical evolution of the corresponding quantum fields $\hat{\Psi}_i({\bf x},t)$, where $i=1,2$ stands for the pulse in a coherent state and a single photon state, respectively.

Let's consider a general interaction between the two light fields in a nonlinear medium.
For the slowly varying and paraxially approximated light fields,
the system evolves with the Hamiltonian $\hat{H}=\hat{K}+\hat{V}$, where
\begin{eqnarray}
\hat{K}=\sum_{i=1}^{2}\int d^3 x\hat{\Psi}_i^{\dagger}({\bf x},t) \{ v_i\frac{1}{i}\nabla_{z}-v_i\frac{\nabla^2_{x}+\nabla^2_{y}}{2k_0}\}\hat{\Psi}_i({\bf x},t)~~
\label{kinetic}
\end{eqnarray}
is the kinetic term \cite{g-c-08}, and the interaction term
\begin{eqnarray}
\hat{V}&=&\int d^3 x\int d^3 x'\hat{\Psi}_1^{\dagger}({\bf x},t)\hat{\Psi}_2^{\dagger}({\bf x}',t)\Delta({\bf x}-{\bf x}')\nonumber\\
&& \hat{\Psi}_2({\bf x}',t)\hat{\Psi}_1({\bf x},t)
\label{interaction}
\end{eqnarray}
describes the general two-body field interaction \cite{fetter}. Here the pulses are assumed to propagate along the $z$ axis with the group velocities $v_i$, and $k_0=2\pi/\lambda_0$ is the central wave
number. For example, $\Delta({\bf x}-{\bf x}')$ can be effectively given as $\chi\delta({\bf x}-{\bf x}')$ for XPM between pulses in media under EIT conditions, where the nonlinear interacting rate $\chi$ is determined by the atomic structures and pulse properties in various systems, see, e.g., \cite{h-h-99, lukin-imamoglu, petrosyan, p-k-02, m-z-03, o-03, r-04, w-s-06}.

The evolution equation of the slowly varying field operators $\hat{\Psi}_i({\bf x},t)$,
\begin{eqnarray}
i\frac{\partial}{\partial t}\hat{\Psi}_i({\bf x},t)=[\hat{\Psi}_i({\bf x},t),\hat{H}],
\end{eqnarray}
read
\begin{eqnarray}
(\frac{\partial}{\partial t}+v_i \frac{\partial}{\partial
z}-iv_i\frac{\nabla^2_T}{2k_0})\hat{\Psi}_i({\bf
x},t)=-i\hat{\alpha}_i({\bf x},t)\hat{\Psi}_i({\bf
x},t),
\label{eqofmotion}
\end{eqnarray}
where $\nabla^2_T=\nabla^2_{x}+\nabla^2_{y}$ and
\begin{eqnarray}
\hat{\alpha}_i({\bf x},t)=\int d^3 x' \Delta({\bf x}-{\bf
x'})\hat{\Psi}^{\dagger}_{3-i}({\bf
x}',t)\hat{\Psi}_{3-i}({\bf x}',t).
\label{alpha}
\end{eqnarray}
The formal solutions to Eq. (\ref{eqofmotion}) are the unitary transformation
\begin{eqnarray}
\hat{\Psi}_{i}({\bf x},t)&=&\hat{U}^{\dagger}(t)\hat{\Psi}_{i}({\bf x})\hat{U}(t)\nonumber\\
&=&\mathbb{T}\exp\{i\int_0^{t}dt'[\frac{v_{i}\nabla^2_{T}}{2k_0}
-\hat{\alpha}_{i}({\bf x}_T,z- v_{i}(t-t'))]\} \nonumber\\
&\times & \hat{\Psi}_{i}({\bf x}_T,z- v_it,0)
\end{eqnarray}
of the fields, where $\hat{U}(t)=\mathbb{T}\exp\{-i\int_0^t dt'\hat{H}(t')\}$. In what follows, we will focus on the unitary evolution of the interacting pulses by neglecting the decoherence effects such as pulse absorption, etc. For XPM between pulses in EIT media, the pulse loss can be neglected given an EIT transparency window much larger than pulse bandwidth. The assumption of a real-number interaction potential $\Delta({\bf x}-{\bf x}')$ is also adopted in the discussions below (i.e. we assume that there is no interaction-induced loss).

\section{interaction picture}
We will use the interaction picture to study the evolution of photonic states. By using this picture one can eliminate the effects of pulse propagation and pulse diffraction, which are due to the kinetic term $\hat{K}$, from the relevant calculations, cf. below.
 The interaction picture is especially helpful to simplify the description of interacting pulses with unequal group velocities $v_i$.

In the interaction picture an operator $\hat{O}({\bf x})$ is transformed to $\hat{O}_I({\bf x},t)=\hat{U}_0^{\dagger}(t)\hat{O}({\bf x})\hat{U}_0(t)$,
where $\hat{U}_0(t)=\exp\{-i\int_0^t dt'\hat{K}(t')\}$.
For example, the field operators in the interaction picture will be
\begin{eqnarray}
\hat{\Psi}_{I,i}({\bf x},t)&=&\exp\{i\int_0^t dt'\hat{K}(t')\}\hat{\Psi}_{i}({\bf x})\exp\{-i\int_0^t dt'\hat{K}(t')\}\nonumber\\
&=&\hat{\Psi}_{i}({\bf x})+[i\int_0^t dt'\hat{K}(t'), \hat{\Psi}_{i}({\bf x})]\nonumber\\
&+&\frac{1}{2!}[i\int_0^t dt'\hat{K}(t'),[i\int_0^t dt'\hat{K}(t'),\hat{\Psi}_{i}({\bf x})]]+\cdots \nonumber\\
&=& \exp\{-v_it\frac{\partial}{\partial z}+iv_it\frac{\nabla^2_T}{2k_0}\}\hat{\Psi}_{i}({\bf x})\nonumber\\
&=&\exp\{iv_it\frac{\nabla^2_T}{2k_0}\}\hat{\Psi}_{i}({\bf x}-v_it\hat{e}_z).
\label{inter-field}
\end{eqnarray}
The interaction Hamiltonian will be correspondingly
\begin{eqnarray}
\hat{V}_I(t)&=&\exp\{i\int_0^t dt'\hat{K}(t')\}\hat{V}\exp\{-i\int_0^t dt'\hat{K}(t')\}\nonumber\\
&=& \int d^3 x_1\int d^3 x_2\hat{\Psi}_{I,1}^{\dagger}({\bf x}_1,t)\hat{\Psi}_{I,2}^{\dagger}({\bf x}_2,t)\Delta({\bf x}_1-{\bf x}_2)\nonumber\\
&&\hat{\Psi}_{I,2}({\bf x}_2,t)\hat{\Psi}_{I,1}({\bf x}_1,t).
\end{eqnarray}
The transverse Laplacians in $\hat{\Psi}_{I,i}$ of (\ref{inter-field}) modify the interaction potential $\Delta({\bf x}_1-{\bf x}_2)$ in the situation of $v_1=v_2=v$ as follows (the expression for the general situation of $v_1\neq v_2$ is similar though more complicated):
\begin{eqnarray}
\Delta'({\bf x})&=&\exp\{-ivt\frac{\nabla^2_{T,x}}{2k_0}\}\Delta({\bf x})\exp\{ivt\frac{\nabla^2_{T,x}}{2k_0}\}\nonumber\\
&=&\Delta({\bf x})-i\frac{vt}{2k_0}\nabla^2_{T,x}\Delta({\bf x})+\cdots,
\label{interplay}
\end{eqnarray}
where ${\bf x}={\bf x}_1-{\bf x}_2$ is the relative coordinate. This effect is discussed in a different way in \cite{trans}. If the propagation length of the pulses is much shorter than their Rayleigh length, the correction terms to $\Delta({\bf x})$ will be insignificant.  Moreover, given a weak interaction where the average of $\hat{\alpha}_i({\bf x},t)$ in Eq. (\ref{alpha}) is a small term, the correction terms from the interplay between the transverse Laplacian and pulse interaction will be even less and can be well neglected in our discussions.

For the state vectors there is the relation
\begin{eqnarray}
|\Phi (t)\rangle_I= \hat{U}_0^{\dagger}(t)|\Phi (t) \rangle
=\exp\{i\int_0^t dt'\hat{K}(t')\}|\Phi (t)\rangle
\end{eqnarray}
between the state vectors $|\Phi(t) \rangle$ in the Schr{\"o}dinger picture and $|\Phi(t)\rangle_I$ in the interaction picture.
The states in the interaction picture evolve according to the equation
\begin{eqnarray}
i\frac{\partial}{\partial t}|\Phi (t)\rangle_I=\hat{V}_I(t)|\Phi (t)\rangle_I.
\end{eqnarray}
An evolved state in the interaction picture is therefore
\begin{eqnarray}
|\Phi (t)\rangle_I&=&\hat{U}_I (t)|\Phi (0)\rangle_I\nonumber\\
&=&\exp\{-i\int_0^t dt'\hat{V}_I(t')\}|\Phi (0)\rangle_I.
\label{out-inter}
\end{eqnarray}

\section{cross-phase modulation between continuous-mode pulses}

\subsection{Single photon pair}

We will study XPM between photonic pulses with the interaction picture introduced in the last section.
For a clearer illustration of the technical steps, we first look at the XPM between two individual photons with the input state as
\begin{eqnarray}
|\Phi_{in}\rangle &=& |1\rangle_1|1\rangle_2\nonumber\\
&=&\int d^3 x_1 f_1({\bf x}_1)\hat{\Psi}_1^{\dagger}({\bf x}_1)\int d^3 x_2 f_2({\bf x}_2)\hat{\Psi}_2^{\dagger}({\bf x}_2)|0\rangle,~~~~~~
\end{eqnarray}
where $f_i({\bf x})= \langle 0|\hat{\Psi}_i({\bf x})|1\rangle$ are the pulse profiles.
The state assumes the same form in the interaction picture because $\hat{\Psi}_{I,i}({\bf x},0)=\hat{\Psi}_i({\bf x})$ at $t=0$.
According to Eq. (\ref{out-inter}), the output state due to the photon-photon interaction will be
\begin{eqnarray}
&&|\Phi_{out}\rangle_I =\hat{U}_I (t)|\Phi_{in}\rangle_I \nonumber\\
&=&\int d^3 x_1 f_1({\bf x}_1)\hat{U}_I (t)\hat{\Psi}_{1}^{\dagger}({\bf x}_1)\hat{U}_I ^{\dagger}(t)\nonumber\\
&\times & \int d^3 x_2 f_2({\bf x}_2)\hat{U}_I (t)\hat{\Psi}_{2}^{\dagger}({\bf x}_2)\hat{U}_I^{\dagger} (t)|0\rangle,
\end{eqnarray}
where we have considered the invariance of the vacuum state $\hat{U}_I^{\dagger} (t)|0\rangle=|0\rangle$.
The transformation of the field operators in the above equation is
\begin{eqnarray}
&&\hat{U}_I (t)\hat{\Psi}_{i}({\bf x})\hat{U}_I^{\dagger} (t)= \hat{\Psi}_{i}({\bf x})-[i\int_0^t dt'\hat{V_I}(t'), \hat{\Psi}_{i}({\bf x})]\nonumber\\
&+&\frac{1}{2!}[i\int_0^t dt'\hat{V_I}(t'),[i\int_0^t dt'\hat{V_I}(t'),\hat{\Psi}_{i}({\bf x})]]+\cdots \nonumber\\
&=&\exp\{i\int_0^t dt'\int d^3 x'\Delta'({\bf x}+v_i t'\hat{e}_z-{\bf x}')\hat{\Psi}_{3-i}^{\dagger}({\bf x}',t')\nonumber\\
&\times &\hat{\Psi}_{3-i}({\bf x}',t')\}
\hat{\Psi}_{I,i}({\bf x})
\equiv  \exp\{i\hat{\varphi}_i({\bf x},t)\}\hat{\Psi}_{i}({\bf x}),
\label{ev-1}
\end{eqnarray}
where $\Delta'({\bf x}-{\bf x}')$ is defined in Eq. (\ref{interplay}).
Here we have used the commutator of the field operator, $[\hat{\Psi}^{\dagger}_{i}({\bf x}_i-v_it'\hat{e}_z), \hat{\Psi}_{i}({\bf x})]=-\delta({\bf x}_i-v_it'\hat{e}_z-{\bf x})$,
in computing the commutators such as
\begin{eqnarray}
&&[\hat{V_I}(t'),\hat{\Psi}_{i}({\bf x})]\nonumber\\
&=&[\int d^3 x_1\int d^3 x_2\hat{\Psi}_{1}^{\dagger}({\bf x}_1-v_1t'\hat{e}_z)\hat{\Psi}_{2}^{\dagger}({\bf x}_2-v_2t'\hat{e}_z)\nonumber\\
&& \Delta'({\bf x}_1-{\bf x}_2)
\hat{\Psi}_{2}({\bf x}_2-v_2t'\hat{e}_z)\hat{\Psi}_{1}({\bf x}_1-v_1t'\hat{e}_z),\nonumber\\
&&\hat{\Psi}_{i}({\bf x})]\nonumber\\
&=&-\int d^3 x'\Delta'({\bf x}+v_i t'\hat{e}_z-{\bf x}')\hat{\Psi}_{3-i}^{\dagger}({\bf x}',t')
\hat{\Psi}_{3-i}({\bf x}',t')\nonumber\\
&\times & \hat{\Psi}_{i}({\bf x}).
\end{eqnarray}
Therefore, we obtain the output state
\begin{eqnarray}
|\Phi_{out}\rangle_I &=&\int d^3 x_1 f_1({\bf x}_1)\hat{\Psi}_{1}^{\dagger}({\bf x}_1)\exp\{-i\hat{\varphi}_1({\bf x}_1,t)\}\nonumber\\
&\times & \int d^3 x_2 f_2({\bf x}_2)\hat{\Psi}_{2}^{\dagger}({\bf x}_2)\exp\{-i\hat{\varphi}_2({\bf x}_2,t\}|0\rangle\nonumber\\
&=&\int d^3 x_1 \int d^3 x_2 \{ f_1({\bf x}_1)  f_2({\bf x}_2)\hat{\Psi}_{1}^{\dagger}({\bf x}_1)\nonumber\\
&\times& \underbrace{\exp(-i\hat{\varphi}_1({\bf x}_1,t))\hat{\Psi}_{2}^{\dagger}({\bf x}_2)\exp(i\hat{\varphi}_1({\bf x}_1,t))}\limits_{\hat{K}({\bf x}_1,{\bf x}_2,t)}\}|0\rangle,\nonumber\\
\label{mid-step}
\end{eqnarray}
considering the relation $\exp\{-i\hat{\varphi}_i({\bf x},t)\}|0\rangle=|0\rangle$.
The operator $\hat{K}({\bf x}_1,{\bf x}_2,t)$ of the under-brace in the above equation can be reduced to the form of a field operator multiplied by a c-number phase:
\begin{eqnarray}
&&e^{-i\int_0^t dt'\int d^3 x'\Delta'({\bf x}_1+v_1 t'\hat{e}_z-{\bf x}')\hat{\Psi}_{2}^{\dagger}\hat{\Psi}_{2}({\bf x}',t')}\hat{\Psi}_{2}^{\dagger}({\bf x}_2) \nonumber\\
&&\times e^{i\int_0^t dt'\int d^3 x'\Delta'({\bf x}_1+v_1 t'\hat{e}_z-{\bf x}')\hat{\Psi}_{2}^{\dagger}\hat{\Psi}_{2}({\bf x}',t')}\nonumber\\
&&=\exp\{-i\int_0^t dt'\Delta'({\bf x}_1-{\bf x}_2+(v_1-v_2)t'\hat{e}_z)\}\hat{\Psi}_{2}^{\dagger}({\bf x}_2)\nonumber\\
&&\equiv \exp\{-i\varphi ({\bf x}_1,{\bf x}_2,t)\}\hat{\Psi}_{2}^{\dagger}({\bf x}_2).
\label{phase}
\end{eqnarray}
The commutator, $[\hat{\Psi}_{2}({\bf x}'-v_2t'\hat{e}_z), \hat{\Psi}_{2}^{\dagger}({\bf x}_2)]=\delta({\bf x}'-v_2t'\hat{e}_z-{\bf x}_2)$, is used in deriving the above result.
Under unitary evolution, the output state of an interacting photon pair therefore reads
\begin{eqnarray}
|\Phi_{out}\rangle_I
&=& \int d^3 x_1 \int d^3 x_2 f_1({\bf x}_1)  f_2({\bf x}_2)e^{-i\varphi ({\bf x}_1,{\bf x}_2,t)}\nonumber\\
&\times & \hat{\Psi}_{1}^{\dagger}({\bf x}_1)\hat{\Psi}_{2}^{\dagger}({\bf x}_2)|0\rangle,
\end{eqnarray}
which is inseparable with respect to ${\bf x}_1$ and ${\bf x}_2$ due to the induced phase $\varphi ({\bf x}_1,{\bf x}_2,t)$. Such inseparability shows the entanglement between the single photon pulses, impairing the ideal performance of a quantum phase gate, $|1\rangle_1|1\rangle_2\rightarrow e^{i\theta_c}|1\rangle_1|1\rangle_2$. The conditional phase $\theta_c$ and fidelity $F$ of a quantum phase gate based on such XPM between two single photons can be determined by the following overlap \cite{trans}:
\begin{eqnarray}
\sqrt{F}e^{-i\theta_c}&=&_I\langle \Phi_{in}|\Phi_{out}\rangle_I \nonumber\\
&=& \int d^3 x_1 \int d^3 x_2 \{f^2_1({\bf x}_1)  f^2_2({\bf x}_2)e^{-i\varphi ({\bf x}_1,{\bf x}_2,t)}\}.\nonumber\\
\label{p-f}
\end{eqnarray}
Obviously a more entangled output state means a lower fidelity $F$ of the gate operation.

The interaction picture adopted in the above discussions makes it unnecessary to consider the pulse propagation and pulse diffraction effects described by Eq. (\ref{kinetic}), which in any case have no impact on the fidelity and conditional phase defined in Eq. (\ref{p-f}), cf. Ref. \cite{trans}. The pulse profiles $f_1({\bf x})$ and $f_2({\bf x})$ in computing the overlap with Eq. (\ref{p-f}) are just those of the initial state at $t=0$. This helps to simplify the treatment of XPM especially when two pulses have unequal velocities $v_1\neq v_2$.

\subsection{Coherent state and single photon state}

We now turn to the main topic of this paper, namely the interaction of a coherent state and a single photon. In the interaction picture, the input of a coherent state and a single photon state is given as
\begin{eqnarray}
|\Phi_{in}\rangle_I &=&|\{\alpha_{\bf k}\}\rangle_I\otimes |1\rangle_I \nonumber\\
&=& \exp \{ \int d^3 x\alpha({\bf x})\hat{\Psi}_{1}^{\dagger}({\bf x})-\int d^3 x\alpha^{\ast}({\bf x})\hat{\Psi}_{1}({\bf x})\} \nonumber\\
&\otimes & \int d^3 y f({\bf y})\hat{\Psi}_{2}^{\dagger}({\bf y})|0\rangle \nonumber\\
&=& \exp\{-\frac{1}{2}\int d^3 x |\alpha({\bf x})|^2 \}\exp \{ \int d^3 x \alpha({\bf x})\hat{\Psi}_{1}^{\dagger}({\bf x})\}\nonumber\\
&\otimes & \int d^3 y f({\bf y})\hat{\Psi}_{2}^{\dagger}({\bf y})|0\rangle.
\label{in-coherent-photon}
\end{eqnarray}
The above input state
will evolve under the interaction $\hat{V}_I$ as
\begin{eqnarray}
&&|\Phi_{out}\rangle_I =\hat{U}_I (t)|\Phi_{in}\rangle_I \nonumber\\
&=& e^{-\frac{1}{2}\int d^3 x |\alpha({\bf x})|^2} \exp \{ \int d^3 x \alpha({\bf x})\hat{\Psi}_{1}^{\dagger}({\bf x})e^{-i\hat{\varphi}_1({\bf x},t)}\}  \nonumber\\
&\times & \int d^3 y f({\bf y})\hat{\Psi}_{2}^{\dagger}({\bf y})e^{-i\hat{\varphi}_2({\bf y},t)}|0\rangle\nonumber\\
&=&e^{-\frac{1}{2}\int d^3 x |\alpha({\bf x})|^2}\sum_n \frac{(\int d^3 x \alpha({\bf x})\hat{\Psi}_{1}^{\dagger}({\bf x})e^{-i\hat{\varphi}_1({\bf x},t)})^n}{n!}\nonumber\\
&\times & \int d^3 y f({\bf y})\hat{\Psi}_{2}^{\dagger}({\bf y})|0\rangle\nonumber\\
&=&e^{-\frac{1}{2}\int d^3 x |\alpha({\bf x})|^2}\sum_n\frac{1}{n!}\{\int d^3 x_1 \alpha({\bf x}_1)e^{-i\varphi({\bf x}_1,{\bf y})}\hat{\Psi}_{1}^{\dagger}({\bf x}_1)\nonumber\\
&& \cdots \int d^3 x_n \alpha({\bf x}_n)e^{-i\varphi({\bf x}_n,{\bf y})}\hat{\Psi}_{1}^{\dagger}({\bf x}_n)\}\nonumber\\
&\times & \int d^3 y f({\bf y})\hat{\Psi}_{2}^{\dagger}({\bf y})|0\rangle\nonumber\\
&=& e^{-\frac{1}{2}\int d^3 x |\alpha({\bf x})|^2}\exp\{\int d^3 x \alpha({\bf x})e^{-i\varphi({\bf x},{\bf y})}\hat{\Psi}_{1}^{\dagger}({\bf x})\}\nonumber\\
&\times & \int d^3 y f({\bf y})\hat{\Psi}_{2}^{\dagger}({\bf y})|0\rangle
\label{out-xpm}
\end{eqnarray}
In the $n$-th term of the above expansion there are $n$ phase factor operators $e^{-i\hat{\varphi}_1}$, see the second step of the equation.
Similar to the procedure in Eq. (\ref{phase}), the commutations of these $n$ phase factor
operators $e^{-i\hat{\varphi}_1}$ with the field operator $\hat{\Psi}_{2}({\bf y})$ of the single photon will give rise to $n$ c-number phase functions $e^{-i\varphi({\bf x}_i,{\bf y})}$, where $\varphi({\bf x}_i,{\bf y})$ is defined in Eq. (\ref{phase}). The last equality in Eq. (\ref{out-xpm}) is one of the main results of this paper. It gives a compact expression for the output state, which forms the basis for all of our results below.

One can see that in general the output state is different from the ideal output state described in the introduction. The output state in Eq. (\ref{out-xpm}) can be expanded as
\begin{eqnarray}
&&e^{-\frac{1}{2}\int d^3 x |\alpha({\bf x})|^2}\{\int d^3 y f({\bf y})\hat{\Psi}_{2}^{\dagger}({\bf y})\}|0\rangle\nonumber\\
&+&\int d^3 x \int d^3 y [\alpha({\bf x})f({\bf y})e^{-i\varphi({\bf x},{\bf y})}\hat{\Psi}_{1}^{\dagger}({\bf x}) \hat{\Psi}_{2}^{\dagger}({\bf y})]|0\rangle\nonumber\\
&+& \mbox {terms involving more than two photons}\}.
\end{eqnarray}
The Fourier transform of the coefficient functions, such as that of the function $\alpha({\bf x})f({\bf y})e^{-i\varphi({\bf x},{\bf y})}$ in the second term of bi-photon component, will be generally inseparable with respect to the wave-vector modes ${\bf k}$ of the coherent state and the modes ${\bf k'}$ of the single photon. This is due to the phases $e^{-i\varphi({\bf x}_i,{\bf y})}$ (inseparable with respect to ${\bf x}_i$ and ${\bf y}$) arising from pulse interaction. However, the XPM for realizing an ideal quantum phase gate is that
$\prod_{\bf k}|\alpha_{\bf k} \rangle\otimes \sum_{{\bf k'}}\xi_{{\bf k'}}a^{\dagger}_{\bf{k'}}|0\rangle
\rightarrow \prod_{\bf k}|\alpha_{\bf k}e^{-i\theta}\rangle\otimes \sum_{{\bf k'}}\xi_{{\bf k'}}a^{\dagger}_{\bf{k'}}|0\rangle$,
i.e. all components $|\alpha_{\bf k}\rangle$ should pick up the same constant phase $\theta$.
Each expanded term in the ideal output state should be still separable with respect to ${\bf k}$ and ${\bf k'}$ as in the input state. The entanglement between the modes ${\bf k}$ and
${\bf k'}$ in a realistic output, as well as the inhomogeneity in pulse interaction, will generally impair such ideal performance for a quantum phase gate.

Another interesting observation is the existence of the self-phase modulation (SPM) effect in a
more general situation with the factor in Eq. (\ref{alpha}) being replaced by
\begin{eqnarray}
\hat{\alpha}_i({\bf x},t)&=&\int d^3 x' \Delta_1({\bf x}-{\bf
x'})\hat{\Psi}^{\dagger}_{i}({\bf
x}',t)\hat{\Psi}_{i}({\bf x}',t)\nonumber\\
&+&\int d^3 x' \Delta_2({\bf x}-{\bf
x'})\hat{\Psi}^{\dagger}_{3-i}({\bf
x}',t)\hat{\Psi}_{3-i}({\bf x}',t). ~~~~~
\label{alpha2}
\end{eqnarray}
Here the potentials $\Delta_1({\bf x}-{\bf
x'})$ and $\Delta_2({\bf x}-{\bf
x'})$ give rise to SPM and XPM, respectively.
The contribution of the SPM effect to the output state in Eq. (\ref{out-xpm}) is manifested by the commutations of $e^{-i\hat{\varphi}_1}$, where
\begin{eqnarray}
&&\hat{\varphi}_1({\bf x},t)=\int_0^t dt'\int d^3 x'\Delta'_1({\bf x}+v_1 t'\hat{e}_z-{\bf x}')(\hat{\Psi}_{1}^{\dagger}\hat{\Psi}_{1})({\bf x}',t')\nonumber\\
&& + \int_0^t dt'\int d^3 x'\Delta'_2({\bf x}+v_1 t'\hat{e}_z-{\bf x}')(\hat{\Psi}_{2}^{\dagger}\hat{\Psi}_{2})({\bf x}',t'),
\end{eqnarray}
with the field operator $\hat{\Psi}_{1}({\bf x})$ of the coherent state, giving the extra phases due to the self interaction of the pulse in coherent state. This is totally different from the interaction between single photons, where only XPM effect exists \cite{trans}.
There is, however, no effect of SPM in multi-photon pulse interaction inside EIT media \cite{eit-review}, and we will not consider SPM below.

Now we look in more detail at the difference of the actual XPM from an ideal XPM which realizes the transformation $\prod_{\bf k}|\alpha_{\bf k} \rangle\otimes |1\rangle \rightarrow \prod_{\bf k}|\alpha_{\bf k}e^{-i\theta}\rangle\otimes |1\rangle$.
In the spatial coordinate an ideal output state is given as
\begin{eqnarray}
|\Phi^0_{out}(\theta)\rangle_I &=&e^{-\frac{1}{2}\int d^3 x |\alpha({\bf x})|^2 }\exp\{ \int d^3 x \alpha({\bf x})e^{-i\theta}\hat{\Psi}_{1}^{\dagger}({\bf x})\}\nonumber\\
&\otimes & \int d^3 y f({\bf y})\hat{\Psi}_{2}^{\dagger}({\bf y})|0\rangle.
\label{ideal-output}
\end{eqnarray}
One can measure the closeness between an actual given by Eq. (\ref{out-xpm}) and an ideal output state via their overlap
\begin{eqnarray}
&& _{I}\langle \Phi^0_{out}(\theta)|\Phi_{out}\rangle_I
=e^{-\int d^3 x |\alpha({\bf x})|^2}\nonumber\\
&\times & \int d^3 y |f({\bf y})|^2 \sum_n\frac{1}{n!}\{\int d^3 x_1 |\alpha({\bf x}_1)|^2 e^{i(\theta-\varphi({\bf x}_1,{\bf y}))} \nonumber\\
&\cdots &\int d^3 x_n |\alpha({\bf x}_n)|^2 e^{i(\theta-\varphi({\bf x}_n,{\bf y}))}\}\nonumber\\
&=& e^{-\bar{n}}\int d^3 y f({\bf y})|^2\sum_n\frac{1}{n!} (\int d^3 x |\alpha({\bf x})|^2 e^{i(\theta-\varphi({\bf x},{\bf y}))})^n\nonumber\\
&=&e^{-\bar{n}}\int d^3 y |f({\bf y})|^2 \exp\{\int d^3 x |\alpha({\bf x})|^2 e^{i(\theta-\varphi({\bf x},{\bf y}))})\}\nonumber\\
&=& e^{-\bar{n}}\int d^3 y |f({\bf y})|^2 \exp\{\int d^3 x |\alpha({\bf x})|^2 \cos(\theta-\varphi({\bf x},{\bf y}))\}\nonumber\\
&\times & \exp\{i\int d^3 x |\alpha({\bf x})|^2 \sin(\theta-\varphi({\bf x},{\bf y}))\},
\label{overlap}
\end{eqnarray}
where $\bar{n}=\int d^3 x |\alpha({\bf x})|^2$ is the average photon number of the coherent state.
In deriving the above result, we have used the relation
\begin{eqnarray}
&&\langle 0|\hat{\Psi}_{}({\bf x'}_n)\cdots\hat{\Psi}_{}({\bf x'}_1)\hat{\Psi}_{}^{\dagger}({\bf x}_1)\cdots \hat{\Psi}_{}^{\dagger}({\bf x}_n)|0\rangle\nonumber\\
&=&\sum_{\mbox{n! terms}}\prod_i \delta({\bf x}_i-{\bf x'}_{P(i)}),
\end{eqnarray}
where the permutations send $i\rightarrow P(i)$, in the first step. It is straightforward to see that, given no interaction between the pulses, i.e., $\varphi({\bf x},{\bf y})=0$ in Eq. (\ref{overlap}), we will have $ _{I}\langle \Phi^0_{out}(0)|\Phi_{out}\rangle_I= _{I}\langle \Phi_{in}|\Phi_{out}\rangle_I=1$.

The fidelity of the XPM between a coherent state and a single photon, which measures how close the actual output is to an ideal output in Eq. (\ref{ideal-output}), can now be defined as
\begin{eqnarray}
F(\theta)&=&|_I\langle\Phi^0_{out}|\Phi_{out}\rangle_I|^2\nonumber\\
&=&|\int d^3y |f({\bf y})|^2\exp \{-\frac{1}{2}\int d^3 x|\alpha({\bf x})e^{-i\theta}\nonumber\\
&-& \alpha({\bf x})e^{i\varphi({\bf x},{\bf y})}|^2 \}|^2.
\label{fidelity-c-s}
\end{eqnarray}
This general formula shows that the fidelity mainly depends on the distance between the functions $\alpha({\bf x})e^{i\varphi({\bf x},{\bf y})}$ and $\alpha({\bf x})e^{-i\theta}$.
Given the $\varphi({\bf x},{\bf y})$ determined from pulse interaction of arbitrary intensity (the nonlinearity due to XPM may not be weak), there exists a $\theta_c$ (up to the modulo of $2\pi$) such that the fidelity $F(\theta)$ is the maximum. We define this $\theta_c$, which corresponds to the ideal output state $|\Phi^0_{out}(\theta_c)\rangle$ that is closest to the actual output state, as the conditional phase of the XPM process.

\section{performance of cross-Kerr nonlinearity}

In this section we study the performance of the XPM between a continuous-mode coherent state and a continuous-mode single photon. In most applications only a small conditional phase $\theta_c$ is necessary for a cross-Kerr nonlinearity between a coherent and single photon state. A suitable weak cross-Kerr nonlinearity is the
core element for realizing all setups proposed in \cite{imoto,munroND,n-m-04, b-s-05, m-05, j-05, k-p-07, h-b, t-k, s-d, l-l,l-h, n-k-k}. Our approach based on the general fidelity function in Eq. (\ref{fidelity-c-s}), however, applies to strong nonlinearity as well.

To begin with, we will only consider the longitudinal degrees of freedom, assuming that the transverse degrees of freedom are well decoupled from the relevant dynamics by properly confining the pulses in the transverse direction.
Such one-dimensional (1-D) treatment of XPM can be found in, e.g. \cite{eit-review,  lukin-imamoglu, petrosyan}.
The pulse interaction potential in these references is given as the 1-D contact
potential $\Delta(z-z')=\chi\delta(z-z')$. Then the fidelity function given in Eq. (\ref{fidelity-c-s}) will be reduced to
\begin{eqnarray}
F(\theta) &= &|e^{-\bar{n}}\int dz' |f(z')|^2 \exp\{\int dz |\alpha(z)|^2 \nonumber\\
&\times & \cos(\theta-\varphi(z,z'))\}|^2
\label{unequal}
\end{eqnarray}
with $\varphi(z,z')=\chi\int_0^t dt'\delta(z-z'+(v_1-v_2)t')$.
Let us first discuss the case
$v=v_1-v_2\neq 0$. For $v>0$, the phase function $\varphi(z,z')$ is constant and equal to $\chi/v$ provided that
$z-z'+vt>0$ and $z-z'<0$ (the integral with respect to $t'$ should be across the point $z-z'+vt'=0$ where the delta function is non-zero) for all relevant values of $z$ and $z'$, i.e. for all values for which $\alpha(z)$ and $f(z')$ are significantly different from zero. In the case of $v<0$, the condition for the constant $\varphi(z,z')=\chi/v$ becomes $z-z'+vt<0$ and $z-z'>0$ instead. Such conditions mean that the two wave packets have to be initially well separated, then one has to pass through the other, and they have to be well separated again in the final state. Under these conditions one
can simply replace $\varphi(z,z')$ by $\chi/v$ in Eq. (\ref{unequal}). It is then clear that $F(\chi/v)=1$. The conditional phase $\theta_c$ is therefore equal to $\chi/v$ in this case, and the usual single-mode description is appropriate. See Fig. 1 for an illustration of this regime with a specific example.

\begin{figure}
\epsfig{file=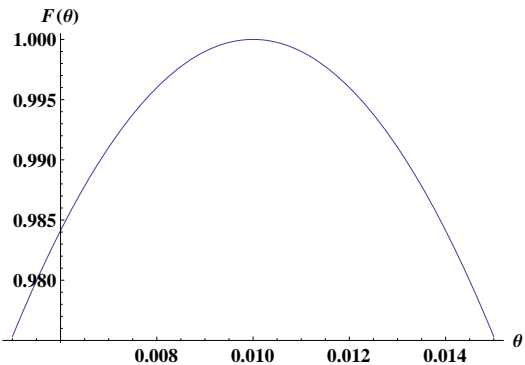,width=0.8\linewidth}
\caption{Fidelity function for the XPM between two pulses moving with a relative velocity $v$. Here the pulses are assumed to have identical Gaussian profiles of width $\sigma$. The initial position of the single photon is $5\sigma$ away from the pulse in coherent state. The parameters for the example are $\bar{n}=10^3$, $\chi/v=0.01$ and $vt=10\sigma$.
The fidelity function assumes the maximum value of 1 at $\theta=\chi/v$.}
\vspace{-0cm} \label{}
\end{figure}

The unit fidelity of the XPM is possible only if one pulse completely passes through the other. The result is
in consistency with Ref. \cite{headon-1, m-10, headon-2}, where the head-on collision or different group velocities of the pulses are proposed to realize a homogeneous phase. Perfect fidelity is possible in principle even for large photon numbers, which is encouraging for potential applications \cite{imoto,munroND, n-m-04, b-s-05, m-05, j-05, k-p-07, h-b, t-k, s-d, l-l,l-h, n-k-k}.

\begin{figure}
\epsfig{file=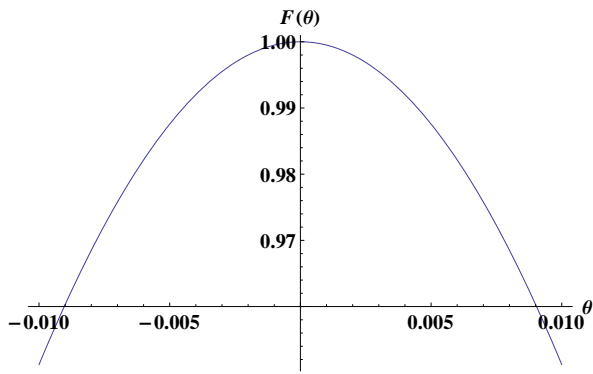,width=0.9\linewidth}
\caption{Fidelity function for the XPM between two co-propagating pulses in the identical Gaussian profile. The parameters for the example are $\bar{n}=10^3$, $\chi t=0.01$. The delta function $\delta(z-z')$ in Eq. (\ref{equal}) is simulated with a sharp Gaussian function $\frac{1}{2\sqrt{\pi\epsilon}}\exp(-\frac{(z-z')^2}{4\epsilon})$, where $\epsilon=10^{-20}$.
The fidelity function assumes the maximum at $\theta=0$.}
\vspace{-0cm} \label{}
\end{figure}

The situation is quite different for co-propagating pulses, i.e. for $v=v_1-v_2=0$. In this case we have
\begin{eqnarray}
F(\theta)&=&|e^{-\bar{n}}\int dz' |f(z')|^2 \nonumber\\
&\times & \exp\{\int dz |\alpha(z)|^2 \cos(\theta-\chi t\delta(z-z')\}|^2.
\label{equal}
\end{eqnarray}
The factor  $\chi t\delta(z-z')$ is non-zero only for $z=z'$, and it appears in the argument of a cosine function, whose modulus is bounded by 1. Therefore it has no effect on the integral over $z$. As a consequence, $F(\theta)$ assumes its maximum value of 1 for $\theta=0$, see Fig. 2 for an illustration by a specific example. As a consequence, the closest ideal output state to the real output state is found to be same as the input state, so the conditional phase $\theta_c$ for this type of XPM is zero. Intuitively this can be understood as follows. The phase shift of the coherent state is due to a single photon. If this photon is detected at a certain position $z$, then the phase of the coherent state will be shifted exactly at that position (i.e. at a single point), but nowhere else. This has a negligible effect when determining the overall phase shift.

Of course, the exact delta potential is only an idealized model. A more realistic model for contact interaction may have very short but non-zero interaction range. Then it will be possible to obtain a small, but non-zero, conditional phase $\theta_c$ for the XPM between two co-propagating pulses. But it is clear that co-propagation is not an attractive regime in the present context.

So far in this section we have only considered the longitudinal degrees of freedom. However, Eq. (\ref{fidelity-c-s}) makes it possible to analyze the effect of the transverse degrees of freedom as well, cf. Ref. \cite{trans}. For simplicity, let us only consider contact interactions, i.e. $\Delta({\bf x-y})=\chi \delta^{(3)}({\bf x-y})$. Let us also assume that the pulses pass through each other as discussed above.
Under these conditions $\varphi({\bf x,y})=\frac{\chi}{v}\delta^{(2)}({\bf x_T-y_T})$, where ${\bf x_T}$ and ${\bf y_T}$ are the transverse components of ${\bf x}$ and ${\bf y}$. The situation is then fully analogous to our above discussion of Eq. (\ref{equal}), where the phase function was also a delta function, and the conditional phase $\theta_c$ is again zero. This simple result holds in the limit where the medium length is much shorter than the Rayleigh length, cf. the discussion for photon-photon gates in Ref. \cite{trans}. This emphasizes the importance of using strong focusing, confinement \cite{headon-2}, or other methods \cite{trans} to suppress these unwanted transverse effects.

\section{conclusions and outlook}

We have developed a general formalism for the continuous-mode treatment of photonic pulse interactions and applied it to the case of the interaction of a coherent state with a single photon, which is highly relevant for applications in quantum information processing. We introduced the concepts of fidelity and conditional phase in order to quantify the validity of the usual single-mode approximations. We found that high fidelities, non-zero conditional phases and high photon numbers (as required for most applications) are compatible in the XPM between two pulses with unequal group velocities, provided that the pulses fully pass through each other. On the other hand, using two exactly co-propagating pulses can hardly generate a non-zero conditional phase. Our results obtained with the general formalism also apply to the regime beyond weak nonlinearity, which is widely considered in the applications.

We believe that our work constitutes significant progress in making the treatment of coherent state - single photon interactions more realistic. There are nevertheless several points where the theory could still be made more realistic, including the inclusion of loss (non-unitary dynamics), non-instantaneous interactions (and the associated noise), and the effect of the forces due to the interaction on the motion of the pulses.

\begin{acknowledgements}
We thank A. I. Lvovsky, A. MacRae, P. M. Leung and J.-M. Wen for helpful discussions. This work was supported by an AI-TF New Faculty Award and an NSERC Discovery Grant.
\end{acknowledgements}

\end{document}